\documentclass[letterpaper,twocolumn,10pt]{article}
\usepackage{usenix2019_v3}
\usepackage{fullpage}
\usepackage{algorithm}
\usepackage{algpseudocode}
\usepackage{amsmath}
\usepackage{balance}
\usepackage{epsfig}
\usepackage{graphicx}
\usepackage{multirow}
\usepackage[normalem]{ulem}
\usepackage{url}
\usepackage{subcaption}
\usepackage{xspace}
\usepackage{times}
\usepackage{enumitem}
\usepackage{color}
\usepackage{flushend}
\setlist{nosep}
\usepackage{titlesec}
\usepackage{datetime}
\usepackage{hyperref}
\usepackage{nth}
\usepackage{array}

\newcommand{\mytilde}{\raise.17ex\hbox{$\scriptstyle\sim$}}
\newcommand{\myparagraph}[1]{\noindent\textbf{#1.}}

\newcommand{\sysname}{\textsc{Kairos}\xspace} 
\newcommand{\EDIT}[1]{#1}

\graphicspath{ {figures/} }

\begin{document}

\date{}

\title{\bf Lightweight Inter-transaction Caching \\ with Precise Clocks and Dynamic Self-invalidation}


\author{                                                                                                Pulkit A. Misra$^\star$\hspace{.05in}                                                                      Srihari Radhakrishnan$^\star$ \\
\vspace{0.1in}
Jeffrey S. Chase$^\star$ \hspace{.05in}                                                                  Johannes Gehrke$^\ddag$ \hspace{.05in}                                                                     Alvin R. Lebeck$^\star$ \\
\vspace{0.1in}
$^\star${\em Duke University} \hspace{.3in}                                                                $^\ddag${\em Microsoft Corporation}
}
\maketitle

\thispagestyle{empty}

\subsection*{Abstract}

Distributed, transactional storage systems scale by sharding data across servers.
However, 
workload-induced hotspots result in contention, 
leading to higher abort rates and performance degradation.

We present \sysname{}, a transactional key-value storage system that leverages client-side inter-transaction caching and sharded transaction validation to balance the dynamic load and alleviate workload-induced hotspots in the system.
\sysname utilizes precise synchronized clocks to implement self-invalidating leases for cache consistency and avoids the overhead and potential hotspots due to maintaining sharing lists or sending invalidations.


Experiments show that inter-transaction caching alone provides 2.35x the throughput of a baseline system with only intra-transaction caching; adding sharded validation further improves the throughput by a factor of 3.1 over baseline. We also show that lease-based caching can operate at a 30\% higher scale while providing 1.46x the throughput of the state-of-the-art explicit invalidation-based caching.

\label{sec:abstract}
\section{Introduction}
\label{sec:intro}
\myparagraph{Motivation}  Transactional (ACID) key-value stores are a common foundation for data center services.  These stores use sharding to spread data across multiple replica groups for scalable throughput and high availability.  They increasingly incorporate low-latency storage media on the servers:  DRAM~\cite{dragojevic2014farm,dragojevic2015no,ousterhout2011the,lee2015implementing}, non-volatile memory (NVM)~\cite{liu2017dudetm}, or solid-state drives (SSDs)~\cite{misra2017enabling}.

This paper proposes a new approach to {\em inter}-transaction caching and concurrency validation for scalable low-latency stores. 
While {\em intra}-transaction caching is trivial with concurrency control for serializable transactions, systems (e.g., Thor~\cite{adya1995efficient}) that support inter-transaction caching typically use {\it explicit invalidation} to keep client caches consistent across transaction boundaries. This approach is similar to network file systems and other client-server storage systems using classical callback leases~\cite{gray1989leases} (see \S\ref{sec:background}). As we show in \S\ref{sec:evaluation}, explicit invalidation introduces substantial cost for high-performance stores with fine-grained concurrency control, such as transactional stores in the data center.  It also complicates the implementation.  As a result, many recent transactional key-value stores do not address inter-transaction caching at all~\cite{corbett2012spanner,thomson2012calvin,dragojevic2014farm,lee2015implementing,ding2015centiman,zhang2015building,misra2017enabling}.

Full support for caching is important for performance, particularly under read-dominated workloads.  Several recent works emphasize the importance of caching for data center services. For example, auto-sharding systems like Slicer~\cite{adya2016slicer} seek to bound the {\it spread} of requests to each data item across application servers, and show substantial improvements to cache effectiveness.  (This idea is a form of locality-aware request distribution~\cite{pai1998locality}.) 
NetCache~\cite{jin2017netcache} embeds caching of hot data into the network to reduce hotspots caused by skewed power-law popularity distributions, which are common in standard workloads~\cite{cooper2010ycsb,atikoglu2012workload}. However, NetCache does not provide transactional semantics.
It is an open question how best to obtain the benefits of client caching with transactions (See \S\ref{sec:related}).

\myparagraph{Contributions} This paper presents \sysname \footnote{\sysname means ``appropriate time'' in Greek.}, 
a transactional key-value store that supports inter-transaction caching {\em without} explicit invalidations. \sysname follows Milana~\cite{misra2017enabling} in using precise synchronized clocks~\cite{ieee2008ptp, corbett2012spanner, geng2018huygens, lee2016dtp} to enable physical time-based consistency integrated with transactional concurrency control, and adds support for inter-transaction caching. Our prototype of \sysname uses Precision Time Protocol (PTP)~\cite{ieee2008ptp}), which provides timestamps with server clock skew $<$ 1 $\mu$s across a data center network.

\sysname is a client-server transaction system that implements transactional serializability using optimistic concurrency control (OCC~\cite{kung1981on}) based on physical clocks, a technique pioneered by Thor~\cite{adya1995efficient}. In client-server OCC systems, a transaction $T$ executes on a single client: when $T$ requests to commit, the servers {\it validate} $T$,
causing a client to abort/restart $T$ if validation detects a conflict. \sysname leverages {\em sharded validation} from Centiman~\cite{ding2015centiman} to decouple transaction validation from the servers, so that validation scales independently of the storage tier. \sysname adapts this sharded validation to support inter-transaction caching (see \S\ref{sec:comparison_with_centiman}), without the cost of explicit invalidation \textbf{(Contribution 1)}.\footnote{It is important to distinguish two similar terms that are independent: {\it validation} refers to a step of optimistic concurrency control that occurs when a client transaction prepares to commit, while {\it invalidation} refers to a server callback to flush a stale value from a client cache.}


Precise synchronized clocks also enable a simple, stateless, time-to-live (TTL) protocol for cache consistency in \sysname.
Storage servers in \sysname hand out leases to cache popular keys in the usual fashion.  We refer to \sysname leases as ``soft'' because the lease manager need not track leases or send  invalidations (callbacks), although it may do so as an optimization for write-heavy keys\footnote{Called {\em tear-off} blocks in a hardware coherence protocol~\cite{lebeck1995dynamic}.}. 
Instead, cache consistency in \sysname is based on low-cost {\it self-invalidation}~\cite{lebeck1995dynamic} when the lease expires \textbf{(Contribution 2)}.
With soft leases, a client may read stale data from its cache with some probability; \sysname uses OCC validation as a fallback to restart any transaction that reads stale data.   {\em The central challenge for this approach is to set lease times to balance the hit ratio with the cost of stale reads.}
\sysname clients use the observed inter-access (read and write) times of popular keys to adapt lease durations {\it dynamically} (see \S\ref{sec:ideal}) for each key to optimize this tradeoff according to an analytical model \textbf{(Contribution 3)}. \EDIT{The classic paper on lease-based consistency~\cite{gray1989leases} suggested adapting lease times based on access parameters and an analytical model, but we are not aware of any work that develops this idea.}

Numerous works use OCC for ACID transactions~\cite{adya1995efficient,dragojevic2014farm,ding2015centiman,zhang2015building,dragojevic2015no,lee2015implementing,chen2016fast,misra2017enabling, yu2018sundial}. Among these, Milana~\cite{misra2017enabling}, Centiman~\cite{ding2015centiman} and Sundial~\cite{yu2018sundial} are most closely related to \sysname. Milana~\cite{misra2017enabling} uses precise clocks to enhance OCC and reduce transaction abort rates, Centiman~\cite{ding2015centiman} performs sharded transaction validation and Sundial~\cite{yu2018sundial} supports inter-transaction caching using logical time-based leases. However, we are not aware of any prior work that performs dynamic self-invalidation of data from client caches or uses an analytical model for calculating lease duration. \S\ref{sec:related} covers the related work.


\myparagraph{Summary of results}
Evaluation of a \sysname prototype under a variant of YCSB workload~\cite{cooper2010ycsb} reveals that inter-transaction caching alone improves throughput by 2.35x relative to a baseline system with only intra-transaction caching; adding sharded validation further improves throughput by a factor of 3.1 over baseline. Furthermore, our evaluation shows that lease-based inter-transaction caching can operate at a 30\% higher scale while providing 1.46x throughput of classical callback leases (explicit invalidation) in workloads with hot keys.

\section{Overview}
\label{sec:background}

This section summarizes client-server transactions and describes how \sysname leverages precise clocks for low-overhead concurrency control and cache consistency.
\label{sec:client-server}
\begin{figure}[t]
\centering
\includegraphics[height=1.3in,keepaspectratio]{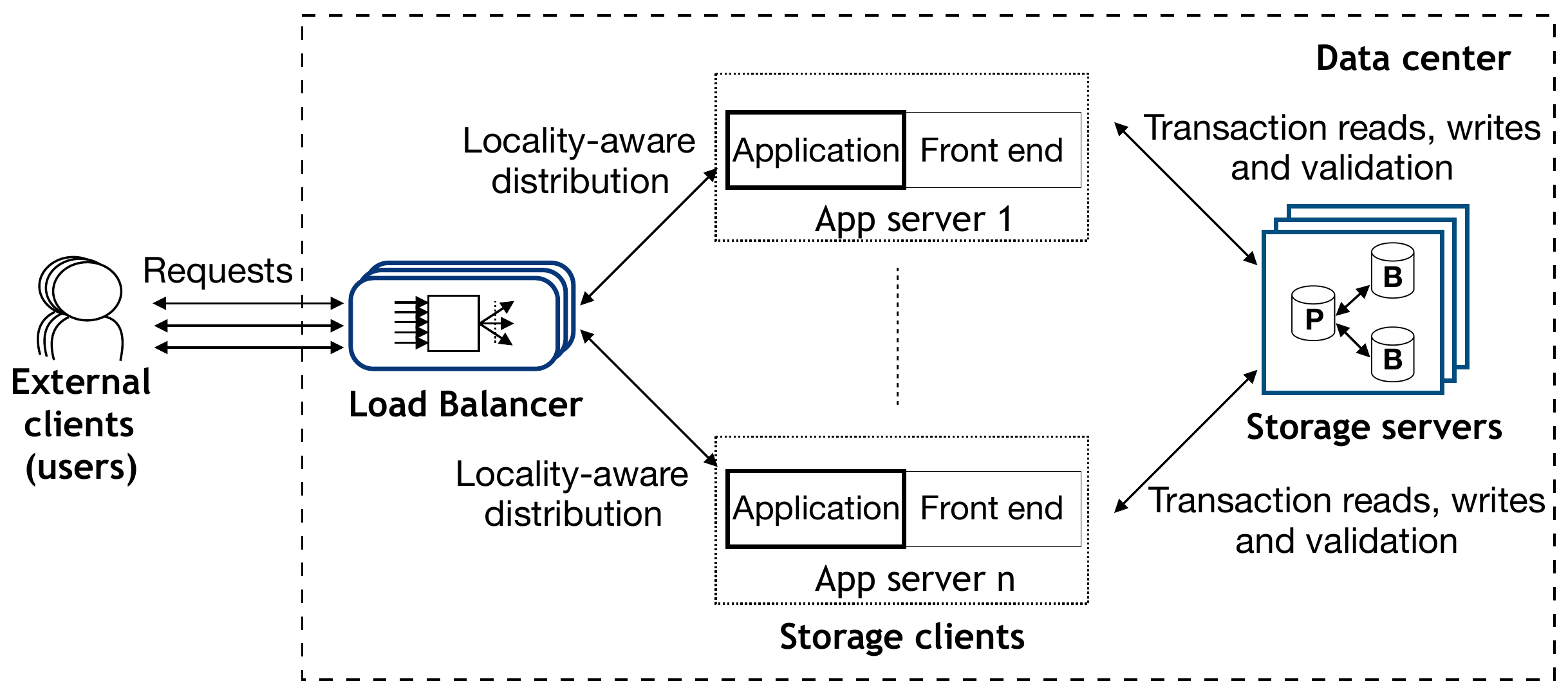}
\caption{Structure of a client-server storage system.}
\label{fig:cs-transactions}
\end{figure}

\myparagraph{Client-server transactions}
Figure~\ref{fig:cs-transactions} depicts a client-server storage system in a data center.  External clients (users) send requests to Web application servers, which run local transactions over stored data, caching objects (key/value data) in their local caches. Load balancers route external requests across the application server tier to balance load and maximize server cache locality. The application servers are the {\it clients} of the storage system; each client has a front-end library to issue transaction operations to the storage servers, manage the local cache, and coordinate local transactions. The key space is partitioned (sharded) across storage server replica groups. 

\sysname uses primary/backup replication, in which one server (the primary) of each replica group handles all incoming read/write requests. The storage clients and primary servers implement OCC by exchanging version timestamps---issued by client physical clocks---for each value read and written.  Each \sysname client (front end) also use these clocks and version timestamps to support {\it inter-transaction} caching.


\myparagraph{Concurrency control}
OCC~\cite{kung1981on,adya1995efficient} ensures serializable ACID transactions. Relative to locking (e.g., two-phase locking or 2PL), OCC enhances concurrency and is not prone to blocking and deadlocks.  OCC systems {\it validate} each transaction $T$ by comparing the version timestamps of $T$'s data values---$T$'s {\it read set} and {\it write set}---to those of other transactions known to the servers when $T$ prepares to commit.
If validation detects a conflict that violates serializable transaction ordering, then $T$'s client aborts and restarts $T$.  It is typical for the primary of each shard to handle validating transactions on the shard.  To reduce the risk that a primary saturates due to validation of hot keys, \sysname adapts {\it sharded validation} from Centiman~\cite{ding2015centiman} to offload this cost to validators, which it co-locates with the clients (see \S\ref{sec:comparison_with_centiman}). 


OCC with physical time leverages precise clocks only for performance: if clocks get out of synch, transactions may get spuriously aborted but no transaction is committed incorrectly, irrespective of the clock skew~\cite{adya1995efficient}. \sysname synchronizes client clocks with the Precision Time Protocol(PTP)~\cite{ieee2008ptp}, which has been shown to reduce abort rates with OCC~\cite{misra2017enabling}.

\label{sec:leases}
\myparagraph{Cache consistency}
In \sysname, {\it precise clocks are also the basis for cache consistency}. Many storage systems implement cache consistency using classical callback leases~\cite{gray1989leases}.  A server $S$ grants a lease to a client $C$ to cache an object $O$ and records the lease.  If $S$ receives a request to update $O$ from another client, it retrieves its record of $C$'s lease, sends $C$ a callback on the lease, and waits for a reply (synchronous) before processing the update.  Each lease is valid for a {\it duration} (term) chosen by the server: a lease specifies a timestamp after which the lease expires.  $C$  considers its cached copy of $O$ to be stale when its lease expires.  In a network file system (e.g., \cite{macklem1994not} or NFSv4) the lease terms may be tens of seconds.

The key observation underlying cache consistency with {\it dynamic self-invalidation} in \sysname is that OCC frees the server from the need to send callbacks.  If the lease term is ``short enough'', then the client marks its copy of $O$ as stale (self-invalidates) before another client updates $O$.  If the lease term is ``too long'', then any client transaction that reads the stale data fails the OCC validation checks and is aborted. The {\it ideal} term is one that allows the lease holder (client) to cache the data long enough to reap some cache hits, and then self-invalidate before it reads stale data. \EDIT{In fact, instead of servers, clients in \sysname adapt the lease terms for popular keys in a dynamic way according to an analytical model that considers the key reference frequency (see \S\ref{sec:caching}). Dynamic self-invalidation offers lightweight cache consistency without the server overhead to calculate lease terms, maintain state records and the network cost and latency of callbacks. The key challenge is for a system to choose lease terms close to the ideal, as measured by the rates of fresh hits and aborts due to stale reads.}


\myparagraph{Precise self-invalidation}
\sysname meets this challenge by using precise clocks to timestamp transaction operations and to set lease terms. With advances in network technology, the one-way network latency ($t_{network}$) is $<$ 10 $\mu$s~\cite{geng2018huygens}.  Storage latencies for stores based on DRAM, NVMs, or SSDs are on similar scales.  Therefore, the inter-access times to objects in a transaction can also be in the $\mu$s range.   Consequently, ideal lease durations may reflect similar time scales. 

In this scenario, clock skew becomes a critical issue for lease-based self-invalidation.
A client with a lagging clock may perceive the lease expiration time as further in the future, so it holds the lease for longer, which increases the probability of reading stale data; similarly, if the client has a leading clock, it expires the lease early, compromising its hit rate.

In the standard Network Time Protocol (NTP), pairs of hosts synchronize their clocks with messages, yielding clock skew  $\epsilon >> t_{network}$ because queuing delays (on servers and within the network) impact the messaging time.   The PTP standard avoids this drawback by assigning timestamps to packets on a server NIC and using ``transparent'' switches which record the ingress and egress time of each clock synchronization packet to account for queuing latencies accurately.  As a result PTP yields $\epsilon \leq t_{network}$.

\section{Inter-transaction caching}
\label{sec:caching}
This section describes lease-based caching (\S\ref{sec:lease_basics}), compares it with other techniques (\S\ref{sec:caching_comparison}) and presents an analytical model to calculate {\em ideal} lease duration (\S\ref{sec:ideal}).


\subsection{Self-invalidation with soft leases}

\label{sec:lease_basics}
\sysname calculates lease duration (term) for a key $K$ based on the observed inter-access (read and write) times of $K$.  We expect that updates to $K$ are independent rather than arriving at regular intervals (although this may occur in some scenarios).  Therefore, the inter-arrival times follow a probability distribution (e.g., exponential) (see \S\ref{sec:ideal}).  Any chosen lease duration for $K$ leaves some probability that an update to $K$ arrives before the lease expires.  In this case, the value is still active on one or more client caches, leaving a window for a client to read a stale value for $K$.  
Stale hits impact forward progress and lead to lower transaction commit rates.

\sysname approximates an {\em ideal} term for each lease (\S\ref{sec:ideal}). The ideal lease term maximizes the expected number of fresh hits and minimizes stale hits. 



\subsection{Comparison of caching techniques}
\label{sec:caching_comparison}
Here we use an example to describe the impact of leases on client cache consistency and set our approach in context with other caching techniques: 1) na\"ive caching, and 2) explicit invalidation (EI).

\begin{figure}[tb]
\centering
 \includegraphics[height=1.5in,keepaspectratio]{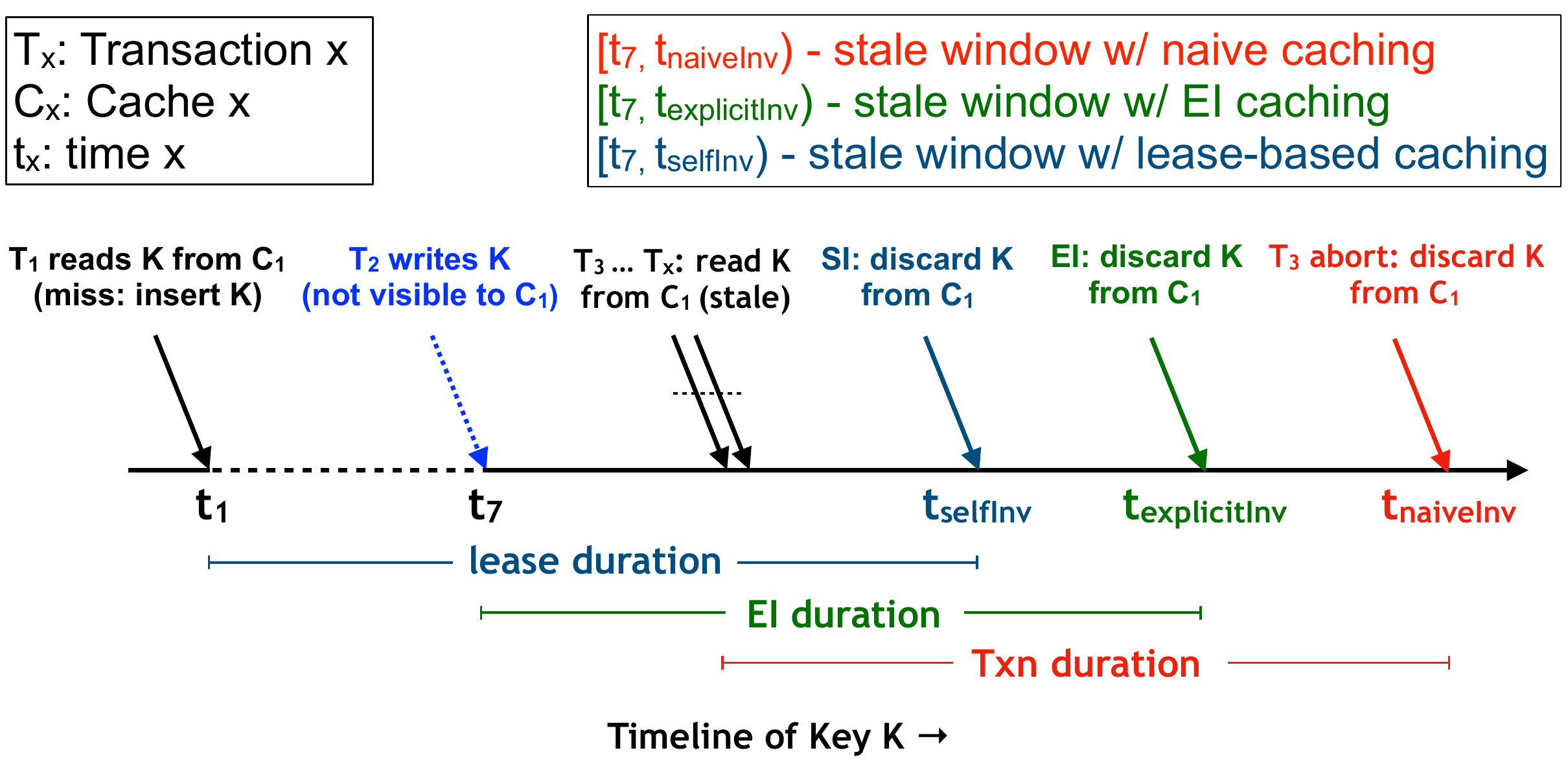} 
 \caption{Impact on client cache consistency with various caching techniques}
\label{fig:timeline}
\end{figure}

\myparagraph{Motivating example} Figure~\ref{fig:timeline} shows the timeline of operations on a key $K$. $K$ is brought into cache $C_1$ at time $t_1$ by transaction $T_1$, which successfully commits soon after (not shown). At time $t_7$, another transaction $T_2$ commits and updates $K$ from a different client (not visible to $C_1$), creating a new version and rendering the cached copy in $C_1$ stale; any transaction that reads $K$ from $C_1$ after $t_7$ reads a stale value and aborts at validation time.

The {\it stale window} is the interval from $t_7$ to the time $C_1$ discards $K$.  This window determines the number of stale hits and the abort rates. Next we consider the stale window with the different caching techniques.

\myparagraph{Na\"ive caching}
Na\"ive caching serves as a baseline straw man. In this approach, a client caches aggressively and discards a cached key only when it learns of a stale read by a local transaction that fails validation. Inter-transaction caching in Sundial~\cite{yu2018sundial} is similar to na\"ive caching (see \S\ref{sec:related}).


Figure~\ref{fig:timeline} shows the consistency challenge with na\"ive caching: a value is determined to be stale only after the first transaction to read the stale value completes ($T_3$ in the figure). In other words, the stale window with na\"ive caching is proportional to the length of a transaction; longer transactions result in more stale hits and higher transaction abort rates.

\myparagraph{Explicit Invalidation (EI)}
In EI caching, servers track {\em sharers} (client caches) that cache a copy of a key $K$ and send an invalidation request (callback) to all sharers on each update to $K$; a sharer cache discards $K$ when it receives the callback. 
Classical leases~\cite{gray1989leases} and Thor~\cite{adya1995efficient} use EI to maintain client cache consistency. 



Figure~\ref{fig:timeline} shows the stale window with EI caching. Key $K$ is updated at time $t_7$ and $C_1$ discards $K$ when it receives the server's callback at $t_{explicitInv}$. Thus, the stale window is [$t_7$, $t_{explicitInv}$). The length of this window is generally the one-way network latency between a server and the client; however, it may be longer if the callback encounters queuing delays. Therefore, any resource constraints (e.g., network bandwidth, CPU cycles) on servers or even the clients can impact the stale window. 


\myparagraph{Lease-based caching}
Figure~\ref{fig:timeline} shows the stale window with self-invalidating leases. Key $K$ is updated at time $t_7$ and $C_1$ discards $K$ when the lease expires at $t_{selfInv}$. 
Lease-based caching does not suffer from the drawbacks of na\"ive caching as the stale window is bounded by the lease end time and is independent of the length of transactions. Moreover, resource constraints do not affect the the stale window nor does the technique require servers to track sharers or send callbacks. Therefore, lease-based caching does not suffer from the drawbacks of EI caching. However, it is sensitive to clock skew.  A leading clock causes the lease to expire sooner and a lagging clock extends the stale window and increases abort risk.

\subsection{Ideal lease duration for a key}
\label{sec:ideal}
The lease duration of a key impacts the stale and overall (fresh + stale) hit rate. Shorter leases incur fewer stale hits but may also reduce the overall cache hit rate. On the other hand, longer leases yield higher overall hit rates but increase the abort rate due to stale hits.

To select the {\em ideal} lease duration, we choose a term that maximizes the expected number of fresh hits.  A more sophisticated solution might consider the weighted cost of transaction aborts due to stale hits in order to choose a suitable level of risk to balance the reward.  We leave that evaluation to future work.

\myparagraph{Arrival rate model for a key} To approximate the ideal lease duration, we need a model of the arrival rate of accesses (read and write) for key $K$. We use a Poisson process to model independent requests for $K$. Poisson is a standard stochastic process for independent arrivals, and is used in popular key-value storage benchmarks (e.g., YCSB~\cite{cooper2010ycsb}, Retwis~\cite{retwis2013}, TPC-C~\cite{leutenegger1993modeling})~\cite{chihoub2012harmony,xiong2015smartsla,pitchumani2015realistic,ren2016igen}. In these benchmarks, each client processes transaction arrivals at a configured {\it transaction arrival rate ($\lambda$)}. Transactions access keys sequentially according to configured logical relationships among keys, e.g., checking the status of an order in TPC-C. However, the decision to access a given key (or set) is independent for each transaction and has no correlation with the access.    The inter-arrival times in a Poisson process are exponentially distributed and the mean inter-arrival time is $\lambda^{-1}$~\cite{trivedi2008probability}. 

Thus $\lambda$ for a given key $K$ depends on its relative {\it popularity}, which is typically modeled as a power law distribution~\cite{cooper2010ycsb,atikoglu2012workload}.  The rate of reads and writes for $K$ depends on its {\it read/write ratio}, which may vary across keys.

\begin{figure}[tb]
\centering
 \includegraphics[height=1.4in,keepaspectratio]{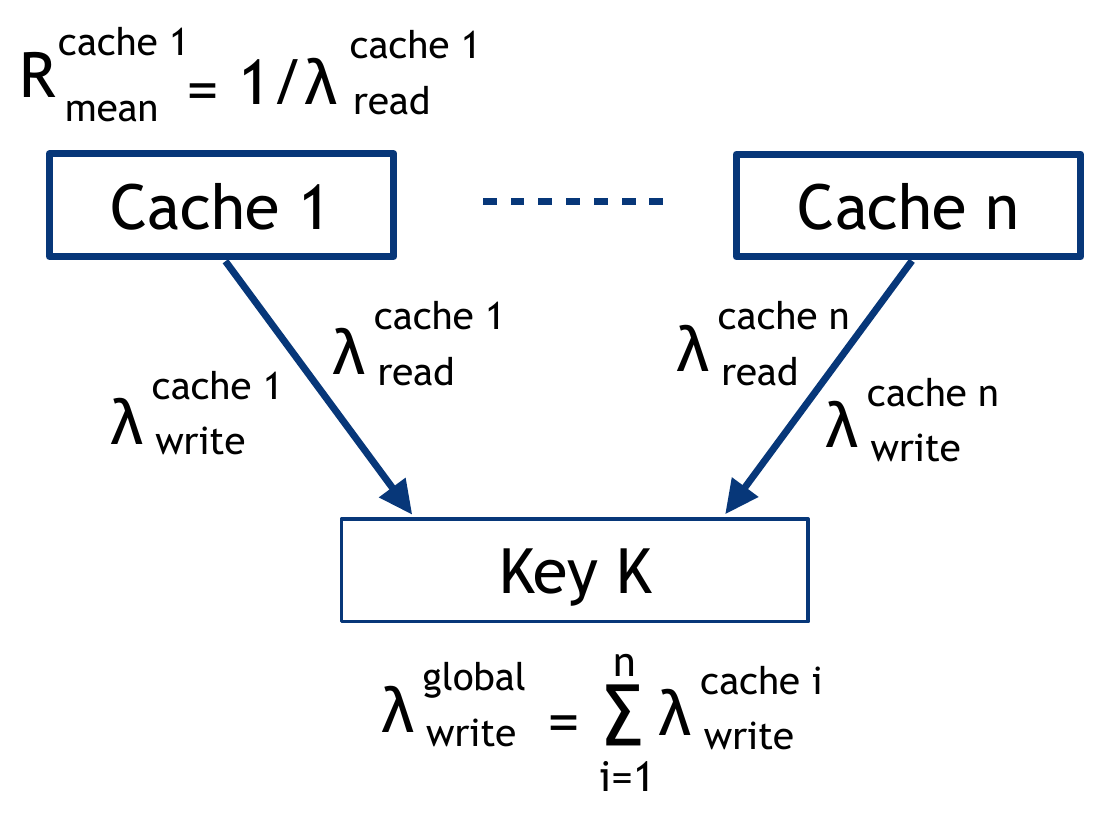} 
 \caption{Arrival rates and inter-access times for a key}
\label{fig:arrival_rates}
\end{figure} 

\myparagraph{Calculating fresh hit rate} In practice, a client (or a server) needs two parameters to evaluate a candidate lease duration for $K$.  Figure~\ref{fig:arrival_rates} illustrates these parameters.  First, the global write arrival rate ($\lambda_{write}^{global}$) of $K$ is needed because a write from any client causes all cached copies to become stale. Second, the per cache mean read inter-arrival time ($R_{mean}^{cache}$) of $K$ allows a server/client to compute the expected {\em per cache} fresh hit rate for $K$.   The global read and write arrival rate for $K$ is the sum of the read and write rates for $K$ across all caches. 

\EDIT{In \sysname, servers track the global mean inter-write time, $W_{mean}^{global}$ (= $\lambda^{-1}$), for the keys they own and each client tracks $R_{mean}^{cache}$ of the frequently-accessed keys. A server reports $W_{mean}^{global}$ for a key $K$ to a client with the response to each GET request for $K$. Using the two values, a client approximates the ideal lease duration for $K$ by generating candidates and evaluating their expected effectiveness.}  Each lease duration $d$ has an expected number of hits within $d$ (given by Equation~\ref{eq:expected_hits}), and a probability of inter-update time being less/greater than $d$.  If W is an exponentially distributed random variable that models the inter-write times of $K$, then the probability that {\em no} update arrives within $d$ is given by Equation~\ref{eq:no_update}.




\begin{align}
\begin{split}
\label{eq:expected_hits}
E[Hits(d)] ={}& \frac{d}{R_{mean}^{cache}}
\end{split}\\
\begin{split}
\label{eq:no_update}
Pr(W > d) ={}& e^{-\lambda_{write}^{global} \times d}
\end{split}\\
\begin{split}
\label{eq:update}
Pr(W \leq d) ={}& 1 - Pr(W > d)\\
={} & 1 - e^{-\lambda_{write}^{global} \times d}
\end{split}
\end{align}

\begin{figure}[tb]
\centering
 \includegraphics[height=1.8in,keepaspectratio]{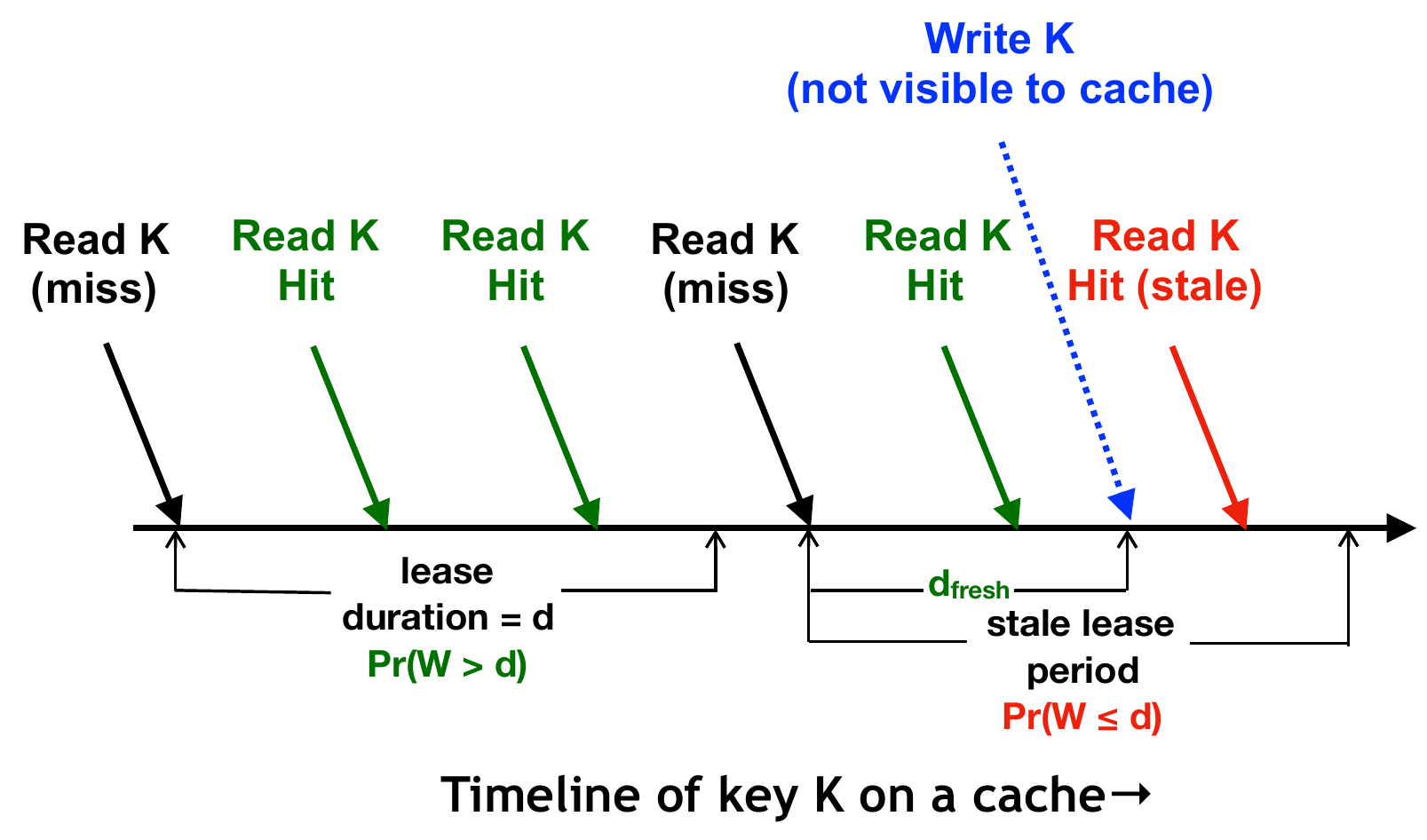} 
 \caption{Fresh hit duration ($d_{fresh}$) in a stale lease period}
\label{fig:stale_period}
\end{figure}

Equation~\ref{eq:update} gives the probability of an update arriving within $d$, i.e., an update arriving while a lease is still active. We call a lease period in which an update arrives a {\it stale lease period}. However, even within a stale lease period, any cache hits that occur before the update are still fresh. Figure~\ref{fig:stale_period} shows the example of a stale lease period and the fresh hit duration ($d_{fresh}$) within the stale lease period. Equation~\ref{eq:expected_fresh} gives the expected value of $d_{fresh}$ in a stale lease period, where $\lambda$ = $\lambda_{write}^{global}$. We use relative times to simplify the equation, i.e., we assume without loss of generality that the stale lease period starts at 0 and ends at $d$.

The fresh hit rate for a lease duration $d$ is the weighted sum of the expected hit rate in a lease period with no update (all hits are fresh) and the expected fresh hit rate in a stale lease period.  A client calculates the fresh hit rate using Equation~\ref{eq:fresh_hit_rate}. Each component is multiplied by the probability that any given lease period is a stale lease period (i.e., an update occurs during the lease term). The denominator is the expected number of cache hits in lease duration $d$ plus the first read miss that fetches the data into the cache. Finally, the expected stale rate for a lease duration $d$ can be calculated by multiplying the hit rate in period ($d$ - $d_{fresh}$) with the probability of a lease period being stale i.e. $Pr(W \leq d)$. 


\begin{align}
\begin{split}
\label{eq:expected_fresh}
E[d_{fresh} | 0 \leq d_{fresh} < d] ={}& \frac{\int_{0}^{d}\lambda x e^{-\lambda x} dx} {\int_{0}^{d} \lambda e^{-\lambda x} dx} \\
={}& \frac{1 - (\lambda d + 1)e^{-\lambda d}}{\lambda (1 - e^{-\lambda d})}
\end{split}
\end{align}

\begin{align}
\begin{split}
\label{eq:fresh_hit_rate}
FreshHitRate(d) ={}& (Pr(W > d) \times HitRate(d)) \\ {}& + (Pr(W \leq d) \times HitRate(d_{fresh})) \\
HitRate(d_{x}) ={}& \frac{E[Hits(d_{x})]}{E[Hits(d)]+1} \text{, $d_x$ = $d$ or $d_{fresh}$}
\end{split}
\end{align}

\begin{figure}[tb]
\centering
 \includegraphics[height=1.8in,keepaspectratio]{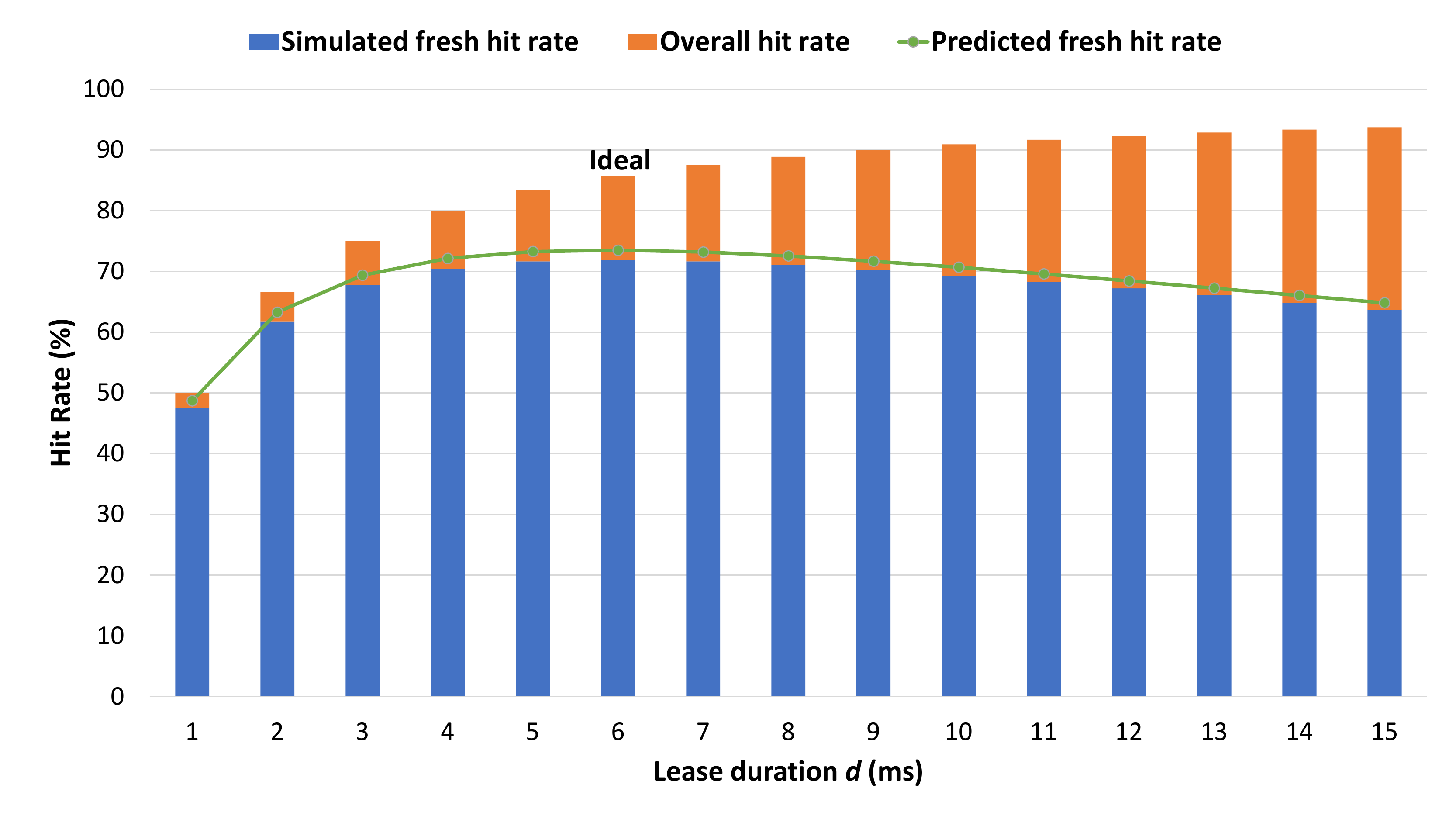} 
 \caption{Ideal lease, $R_{mean}^{cache}$ = 1 ms, $W_{mean}^{global}$ = 19 ms.}
\label{fig:ideal}
\end{figure}

\myparagraph{Finding ideal lease duration} To illustrate, Figure~\ref{fig:ideal} shows the fresh hit rate for varying lease durations for a key with $R_{mean}^{cache}$ = 1 ms and $W_{mean}^{global}$ = 19 ms. It also shows prediction accuracy by comparing predictions using Equation~\ref{eq:fresh_hit_rate} with  results from a Monte Carlo simulation. The simulator takes $R_{mean}^{cache}$, $W_{mean}^{global}$ and a candidate lease duration as the input, and generates read and write arrivals from an exponential distribution with the specified inter-arrival times. Data is cached for the lease duration after each miss; the read arrival times are used to determine if a read is a cache hit, and write arrival times are used to determine whether a hit is fresh. We simulate 10M accesses for each lease duration. As seen from the figure, the predicted values are always optimistic, and the average difference with the simulation results is 1.4\%.

The trend from figure~\ref{fig:ideal} is that the overall hit rate (fresh + stale hits) increases with the lease duration ($d$); the fresh hit rate increases initially, hits a peak, which is the ideal value of $d$, and any subsequent increase in $d$ only increases the number of stale hits and therefore the fresh hit rate starts to drop. We use this trend to design a simple gradient algorithm to find the ideal lease duration for a key $K$ (Algorithm~\ref{alg:find_ideal}). The algorithm takes $R_{mean}^{cache}$ and $\lambda_{write}^{global}$ for $K$ and returns the ideal lease duration and highest fresh hit rate. Other possible considerations (e.g., minimum fresh hit rate, maximum stale rate) are left to future work. \EDIT{As \sysname clients calculate leases independently, each client can use a custom lease calculation technique based on the degree of staleness that it is willing to tolerate.} 


\begin{figure}[tb]
    \begin{algorithm}[H]
    \begin{algorithmic}[1]
  	\scriptsize
    \Procedure {find\_ideal\_lease\_duration}{$R_{mean}^{cache}$, $\lambda_{write}^{global}$}
    	\State bestLeaseDuration = 0
        \State bestFreshHitRate = 0
        \State expectedHitsPerLease = 1
        \State foundIdealLease = false
        \While{!foundIdealLease}
        	\State leaseDuration = expectedHitsPerLease $\times$ $R_{mean}^{cache}$
            \State freshHitRate = fresh\_hit\_rate(leaseDuration, $R_{mean}^{cache}$, $\lambda_{write}^{global}$)
            \If{freshHitRate $<$ bestFreshHitRate}
                \State foundIdealLease = true \Comment{Stop searching}
            \Else
                \State bestFreshHitRate = freshHitRate
                \State bestLeaseDuration = leaseDuration
                \State expectedHitsPerLease += 1 \Comment{Increase lease duration}
            \EndIf
        \EndWhile
        \State return \{bestLeaseDuration, bestFreshHitRate\}
    \EndProcedure
	\caption{Find ideal lease duration}
    \label{alg:find_ideal}
    \end{algorithmic}
    \end{algorithm}
\end{figure}


\section{\sysname: A Transactional Key-Value Storage System}
\label{sec:architecture}


\subsection{System architecture}

\begin{figure}[t]
\centering
\includegraphics[height=1.9in,keepaspectratio]{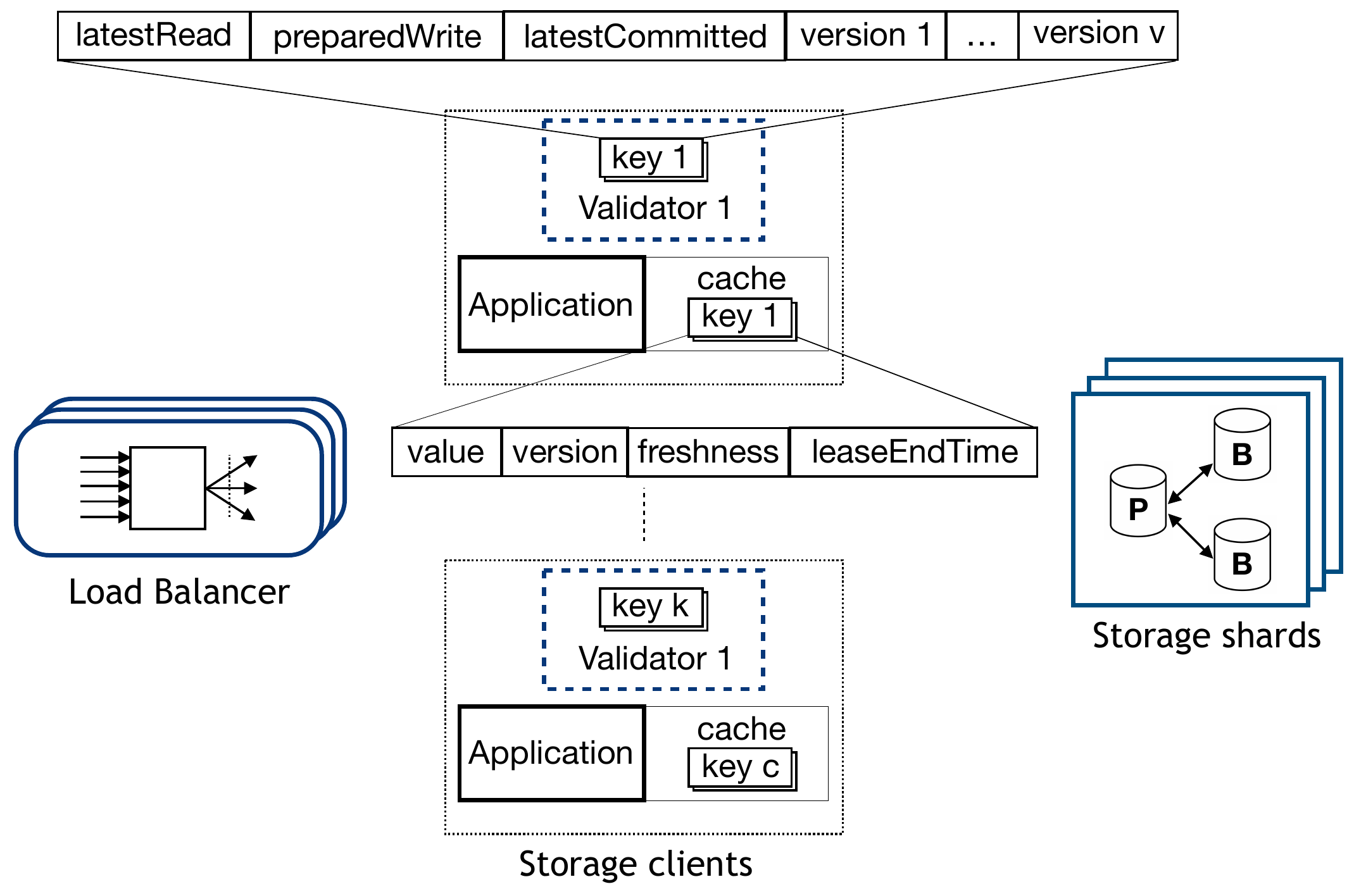}
\caption{\sysname architecture}
\label{fig:architecture}
\end{figure}


Figure~\ref{fig:architecture} shows the architecture of \sysname. \sysname targets the intra-data center client-server storage model described in \S\ref{sec:client-server}. To improve performance, \sysname allows inter-transaction caching on the clients and shards validation workload across client-side validators. 


Each client cache in \sysname operates independently and popular data may be replicated across caches. Each key in a client cache is associated with a value, version timestamp ($ts_{version}$), lease end time ($ts_{leaseEndTime}$) and freshness timestamp ($ts_{freshness}$). $ts_{version}$ is the commit timestamp of the transaction that wrote the version, $ts_{leaseEndTime}$ indicates the time at which the entry will be self-invalidated from the cache and $ts_{freshness}$ indicates the latest timestamp for which the client knows that the cached version is fresh i.e., there are no superseding writes with timestamp $t$ in the interval ($ts_{version}$, $ts_{freshness}$] (see \S\ref{sec:protocol}).


Key ownerships are distributed across client-side validators. A validator maintains a latest read timestamp ($ts_{latestRead}$), a $preparedWrite$ flag, a latest committed timestamp ($ts_{latestCommitted}$) and a list of version timestamps for each key that it owns and is responsible for validating. $ts_{latestRead}$ indicates the highest commit timestamp of a transaction that read the key, $preparedWrite$ indicates if there is a successfully validated but uncommitted transaction and $ts_{latestCommitted}$ is the commit timestamp of the last transaction that wrote the key.

In addition, each validator maintains a garbage collection threshold timestamp $ts_{GC}$ and discards all key versions older than $ts_{GC}$. $ts_{GC}$ indicates that the validator has sufficient state to validate all transactions with freshness timestamps $\geq$ $ts_{GC}$ (see \S\ref{sec:version_management}). 

\subsection{Transaction protocol}
\label{sec:protocol}


\begin{figure}[t]
    \begin{algorithm}[H]
    \begin{algorithmic}[1]
  	\scriptsize
    \Procedure {read}{T, key}
    	\If {key $\in$ T.writeSet}
        	\State return T.writeSet[key].value
        \ElsIf {key $\in$ T.readSet}
        	\State return T.readSet[key].value
        \Else
        	\If {key $\notin$ Cache \textbf{or} Cache[key].lts $<$ $t_{current}$}
            	\State Cache[key].\{value, vts\} = get\_from\_server(key)
            	\State Cache[key].lts = $t_{current}$ + ideal\_lease($R_{mean}^{cache}$, $\lambda_{write}^{global}$)
                \State Cache[key].fts = max(Cache[key].vts, $ts^{global}_{watermark}$)
            \EndIf
            \State T.readSet[key].\{value, vts, fts\} = Cache[key].\{value, vts, fts\}
            \State return T.readSet[key].value
        \EndIf
    \EndProcedure
    \Statex
	\Procedure {write}{T, key, value}
    \State T.writeSet[key].value = value
    \EndProcedure
	\caption{Processing phase of a transaction $T$. \small{For brevity, we use $ts_{version}$ = vts, $ts_{leaseEndTime}$ = lts, $ts_{freshness}$ = fts}}
    \label{alg:execute}
    \end{algorithmic}
    \end{algorithm}
\end{figure}

\sysname executes transactions in a similar manner to other client-server storage systems with OCC~\cite{adya1995efficient, zhang2015building,ding2015centiman}. Algorithm~\ref{alg:execute} shows how a transaction $T$'s reads and write requests are handled during the processing phase of $T$. For a read request, a value is returned if the key exists in $T$'s write or read set. Otherwise, the transaction checks the local cache for the key. \EDIT{On a cache miss, the read request is sent to the remote server; the server returns the value, $ts_{version}$ and $W^{global}_{mean}$. A client calculates the ideal lease duration ($d_{ideal}$) for the key using the cache-observed $R^{cache}_{mean}$ and server-received $W^{global}_{mean}$, and sets $key.lts = t_{current} + d_{ideal}$.} The $ts_{freshness}$ value of a cached key is set to the max of the timestamp of the returned version and the client's view of the global watermark, i.e., $ts_{freshness} = max(ts_{version}, ts^{global}_{watermark})$; \sysname guarantees that no writes older than  $ts^{global}_{watermark}$ will occur in the system (\S\ref{sec:version_management} describes watermarks in greater detail). 
For each key read in the transaction, the read set tracks its value, $ts_{version}$ and $ts_{freshness}$. All transaction writes during the processing phase are buffered and made visible only after a transaction commits, as is typical in OCC. 

The processing phase finishes when the application invokes commit transaction. Before initiating the commit process, the client assigns a freshness and commit timestamp to the transaction $T$. The freshness timestamp of a transaction ($T_{freshness}$) is the minimum freshness timestamp from all keys in the read set, i.e., $T_{freshness} = \displaystyle \min_{key \in readSet} key.fts$. The commit timestamp ($T_{commit}$) depends on the type of transaction (read-only or read-write); for a read-only transaction, the commit timestamp is max freshness timestamp from all keys in the read set i.e., $T_{commit} = \displaystyle \max_{key \in readSet} key.fts$ and for a read-write transaction, $T_{commit} = t_{current}$, where $t_{current}$ is the current time on the client. After assigning the timestamps, the client initiates and acts as the coordinator in the the two-phase commit (2PC) protocol; the 2PC participants include the storage servers and validators for all keys in either set.
The validation decisions of transactions are logged on the client.

\subsection{Validation}
\label{sec:validation}



Algorithm~\ref{alg:validation} shows the validation algorithm used by the sharded client-side validators in \sysname. A validator simply aborts a transaction $T$ if $T_{freshness}$ $<$ $ts_{GC}$ because it does not have the state to validate $T$ as it discards versions behind $ts_{GC}$. $T_{freshness}$ provides a time bound on the oldest key read by $T$ from its client cache; $T$ can be validated only if $T_{freshness} \geq ts_{GC}$ because if T read any stale data from its cache, i.e., it missed a superseding write w, then w's timestamp must be later than $T_{freshness}$, and if $T_{freshness} \geq ts_{GC}$ then the validator must remember w, and the validator aborts T.

Validation of read-only transactions is same as in other OCC-based systems~\cite{adya1995efficient, zhang2015building,ding2015centiman,lee2015implementing}, below we describe our protocol for validating read-write transactions.


\begin{figure}[tb]
\centering
 \includegraphics[height=1.9in,keepaspectratio]{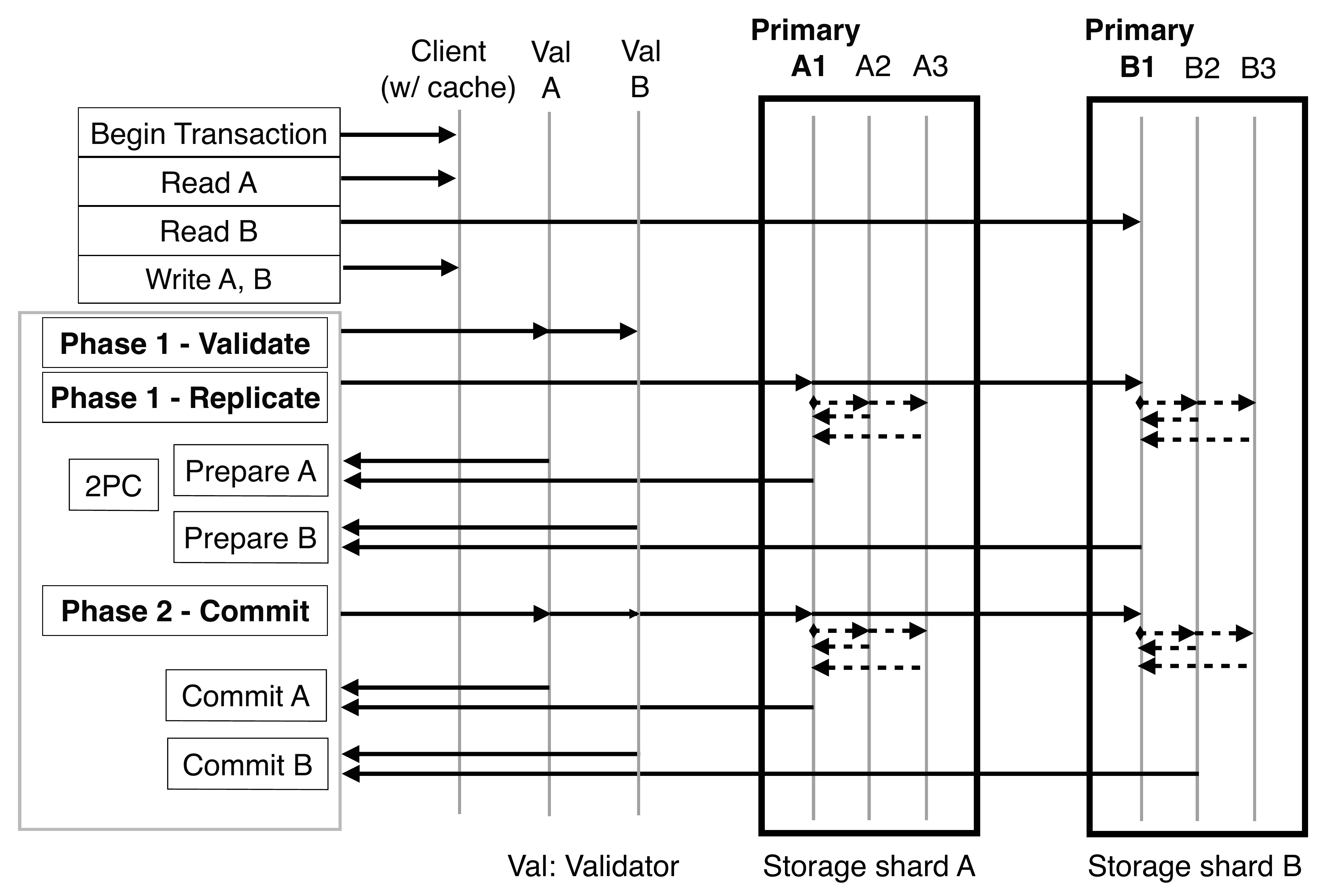} 
 \caption{Two Phase Commit (2PC)}
\label{fig:2PC}
\end{figure}

\begin{figure}[t]
    \begin{algorithm}[H]
      \begin{algorithmic}[1]
       \scriptsize
       \Procedure {validate}{txn}
       
       \If {txn.freshness $<$ $ts_{GC}$}
       \State return ABORT \Comment{Not enough state to validate}
       \EndIf
         \For {each (key, version) $\in$ txn.readSet}
          \If {key.preparedWrite}
              \State return ABORT
           \ElsIf {key.latestCommitted $\neq$ version}
           	  \State return ABORT
           \EndIf
         \EndFor
         \State newVersion = txn.commitTimestamp
         \For {each (key, version) $\in$ txn.writeSet} 
          \If {key.preparedWrite}
              \State return ABORT 
           \ElsIf {key.latestRead $\geq$ newVersion}
           	  \State return ABORT
           \ElsIf {key.latestCommitted $\geq$ newVersion}
              \State return ABORT
           \EndIf
         \EndFor
         \State return COMMIT
          \EndProcedure
        \caption{Validation Algorithm}
        \label{alg:validation}
        \end{algorithmic}
        \end{algorithm}
\end{figure}

\myparagraph{Read-write transaction} Figure~\ref{fig:2PC} shows an example of the distributed two phase commit (2PC) protocol used to commit read-write transactions, with the client acting as the coordinator. The key difference is that in
phase 1 of 2PC, a client (coordinator) sends, in parallel, a validate request to the participant validators and a replicate request to primaries of participant storage shards. Replicate requests are quorum replicated before a primary of a storage shard returns a response. A client accumulates all validate and replicate responses before starting phase 2 of 2PC. A transaction decision is COMMIT only if {\em all} validators respond with a COMMIT during phase 1, otherwise the decision is to ABORT. In phase 2, a client informs the transaction decision to the application and all participant validators and primaries of storage shards. 

Although not shown in the algorithm, after successful validation of a read-only or read-write transaction, a validator sets $key.latestRead$ = $txn.commitTimestamp$ for all validated keys in the read set and $key.preparedWrite$ = $true$ for all keys in the write set of the transaction. A validator resets $key.preparedWrite$ on receiving the transaction decision during 2nd phase of 2PC and also updates $key.latestCommitted$ for a COMMIT decision.


\subsection{Watermarks and version management}
\label{sec:version_management}

Each \sysname client calculates the global watermark ($ts^{global}_{watermark}$) as described in Centiman~\cite{ding2015centiman}; the meaning of $ts^{global}_{watermark}$ is that any transaction in the system with commit timestamp $t$ $\leq$ $ts^{global}_{watermark}$ has already completed. Clients use $ts^{global}_{watermark}$ to assign a freshness timestamp to newly cached keys (see Algorithm~\ref{alg:execute}). Let $c_{freshness}$ be the minimum freshness timestamp from all cached keys in a client. The meaning of $c_{freshness}$ is that the oldest cached value in a client is known to be fresh as of $c_{freshness}$. In other words, the client may have missed writes to cached keys with timestamp $>$ $c_{freshness}$, but knows about all writes with timestamps $<$ $c_{freshness}$. $c_{freshness}$ depends on lease durations and with long leases client caches can have $c_{freshness} << ts^{global}_{watermark}$.

Each client periodically broadcasts $c_{freshness}$ to all storage servers and validators in the system. In turn, the validators and  storage servers use the individual $c_{freshness}$ value to calculate a garbage collection timestamp $ts_{GC}$, where $ts_{GC}$ = $\displaystyle \min_{c \in C}$ $c_{freshness}$. Versions with timestamps $<$ $ts_{GC}$ are discarded by the storage servers and validators. 


\subsection{Comparison with Centiman}
\label{sec:comparison_with_centiman}
There are some similarities between \sysname and Centiman~\cite{ding2015centiman}. \sysname follows Centiman in decoupling validation from storage servers, so that validation
scales independently of the storage tier. Like Centiman, \sysname also maintains multiple versions of (key, version) pairs on the sharded validators for providing transactional serializability. Finally, both \sysname and Centiman use watermarks for version management; watermarks are used for calculating a garbage collection timestamp ($ts_{GC}$) and versions with timestamp $< ts_{GC}$ are discarded.

However, Centiman keeps $ts_{GC}$ $\approx$ $ts^{global}_{watermark}$ (global watermark) to minimize state on validators as transactions always read fresh values from storage servers during execution since there is no inter-transaction caching. In contrast, \sysname aims at keeping $ts_{GC} << ts^{global}_{watermark}$ to maximize state on validators, which in turn enables inter-transaction caching on the clients. In \sysname, a validator cannot validate a transaction $T$ if $T.freshness < ts_{GC}$ (see \S\ref{sec:validation}) because it does not know of all writes that are pertinent to $T$, that $T$ might have missed due to stale cache reads, since the validator discards versions behind $ts_{GC}$. Therefore, keeping $ts_{GC} << ts^{global}_{watermark}$ enables \sysname to validate transactions that read ``old'' data out of the cache. Crucially, it also enables \sysname to use validation as a ``fallback'' for cache consistency: it is safe for $T$ to read stale data without violating consistency because a validator will abort $T$.
\section{Evaluation}
\label{sec:evaluation}
This section presents the results from our prototype implementation of \sysname.

\myparagraph{Implementation} We implemented three client-side ca-\\ching
techniques on top of an existing transactional key-value storage system~\cite{misra2017enabling}: na\"ive, explicit invalidation (EI) and lease-based client-side caching. The involvement of a server in the storage system for maintaining client cache consistency varies, depending on the caching technique. In na\"ive caching, a server is unaware of client-side caching and cache management occurs solely based on transaction decisions. In contrast, in EI caching, a server tracks the sharers (client caches) of a key $K$ and sends invalidations to all the sharers after a read-write transaction that modifies $K$ commits. Finally, in lease-based caching, a server only calculates the global mean inter-write times ($W_{mean}^{global}$) of keys and piggybacks the $W_{mean}^{global}$ of a key $K$ with the response to a GET request for $K$. The ideal lease duration for $K$ is calculated on the client from $W_{mean}^{global}$ and the client-observed mean inter-read time ($R_{mean}^{cache}$) of $K$ using Algorithm~\ref{alg:find_ideal}. 

To compare the performance impact of caching in isolation, we fix the set of keys that can be cached in the three caching techniques to top 1\% keys based on popularity; the keys are pre-computed using a top-k algorithm~\cite{charikar2002finding,metwally2005topk}. This step is particularly important for EI caching because we found that a server saturates in our setup when tasked with maintaining client cache consistency for {\em all} keys that it owns. Consequently, clients can only cache a subset of the top 1\% keys, depending on the cache size. By default, clients cache 0.1\% (of the top keys) with LRU replacement. Furthermore, the servers only track sharers for the top 1\% keys in EI caching. However, for lease-based caching, the servers track $W_{mean}^{global}$ of {\em all} keys, clients track $R_{mean}^{cache}$ of the top 1\% keys and calculate the lease duration of a key after each fetch from the server. Without pre-computed popular keys, memory-efficient techniques proposed in a prior work~\cite{li2016switchkv} can be used to determine popular keys for which clients need to track the mean inter-read time.

For sharded validation, we implement a client-side validator and make the necessary changes on servers for the modified 2PC protocol. Client watermarks and freshness timestamps are updated after every 10k transactions.

\myparagraph{Experimental setup} We run all experiments on Microsoft Azure D4s.v3 nodes with 4 vCPUs, 16 GB of RAM and a high performance network. All experiments use 5 storage shards and up to 30 clients. Each shard has 1 primary and 2 backup servers; data is stored on DRAM on all servers. We co-locate the application and validators on a subset of the clients.
The system clocks on all VMs are synchronized using PTP software timestamping mode. The average clock skew between VMs and one-way network latency is 400 $\mu$s and 500 $\mu$s, respectively.

\myparagraph{Workload} We use a variant of YCSB~\cite{cooper2010ycsb} to model a social network application where the data of popular users is read more often and users have different rate of posting updates. The workload models this behavior by using different zipfian distribution coefficients for controlling popularity of keys in read-only ($\alpha_{r}$) and read-write ($\alpha_{rw}$) transactions. By default, $\alpha_{r}$ = 0.99, $\alpha_{rw}$ = 0.5 and 90\% of transactions are read-only. We evaluate the impact of varying $\alpha_{r}$, $\alpha_{rw}$ and the percentage of read-only transactions. Each transaction accesses 4 keys. We populate the system with 20M keys; each key is 16B in size and a value is 1KB.

\subsection{Performance impact of the different techniques}
Here we compare the performance impact of the two techniques presented in the paper --- inter-transaction caching (explicit invalidation and ideal leases) and client-side validation (CV) --- while varying the read-only transaction ratio and popularity of keys in a workload. Our baseline system performs {\em intra}-transaction caching and server-side validation of all transactions. To this baseline system, we add each technique individually (e.g., inter-transaction caching only, without client-side validation) and evaluate the impact of the addition on performance. Furthermore, we also evaluate the impact of the combination of the two techniques. 

We use 10 clients and 5 storage shards for this set of experiments and vary the offered load per client. Each client caches 0.1\% of the most popular keys in the workload. Finally, we use 5 validators (same as the number of storage shards) in the system configuration with client-side validation of transactions. 

\begin{figure}[t]
\centering
 \includegraphics[width=0.47\textwidth,keepaspectratio]{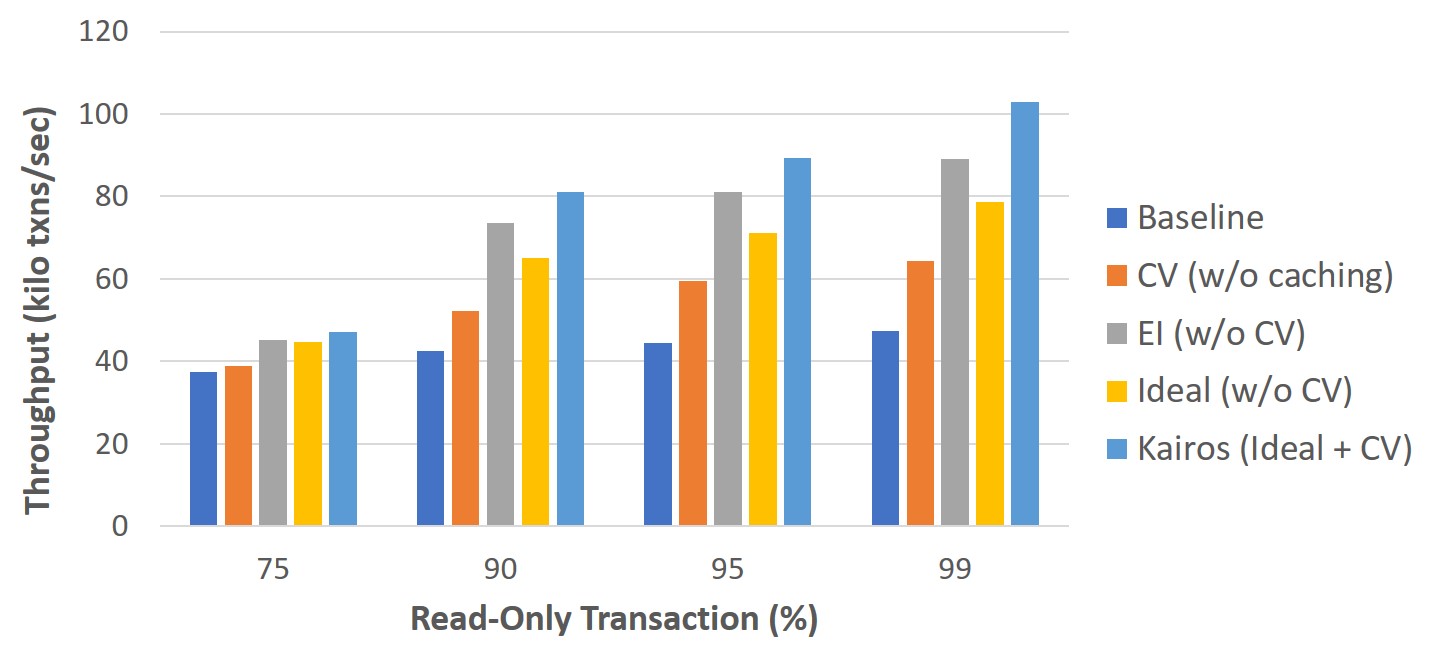}
\caption{Ratio of read-only transactions vs throughput}
\label{fig:txn_rw}
\end{figure}

Figure~\ref{fig:txn_rw} shows the throughput for different ratios of read-only transactions for the various configurations of the system. The figure shows 5 different configurations: 1) baseline, 2) client-side validation (CV) only (no inter-transaction caching), 3) explicit invalidation-based (EI) caching (no CV), 4) ideal lease-based caching (no CV) and 5) \sysname (ideal lease-based caching and CV). 

As seen from the figure, the throughput of all configurations increases with an increase in the ratio of read-only (vs read-write) transactions. For the baseline, the throughput increases because the version management (creating new and discarding old versions) overhead decreases when the ratio of read-only transactions increases. Adding CV only to the baseline system offers up to 42\% higher throughput. The throughput with CV increases with an increase in read-only transaction ratio because the load on the servers to commit a transaction decreases. Recall that servers are involved in the 2PC protocol for all read-write transactions (see Section~\ref{sec:validation}). Adding inter-transaction caching to the baseline system offers up to 2$\times$ improvement in throughput, with EI providing up to 15\% higher throughput than ideal lease-based caching. This is because ideal has a lower cache hit rate (6\% lower on average) per client since it relies on probabilistic leases, which causes it to preemptively discard a cached entry to avoid stale hits; we present a detailed comparison between EI and ideal in Section~\ref{sec:caching_eval}. Finally, a combination of CV and inter-transaction caching (with ideal) offers the highest improvement in performance - up to 2.42$\times$ over baseline.

\begin{figure}[t]
\begin{subfigure}{0.47\textwidth}
    \centering \includegraphics[width=\linewidth,keepaspectratio]{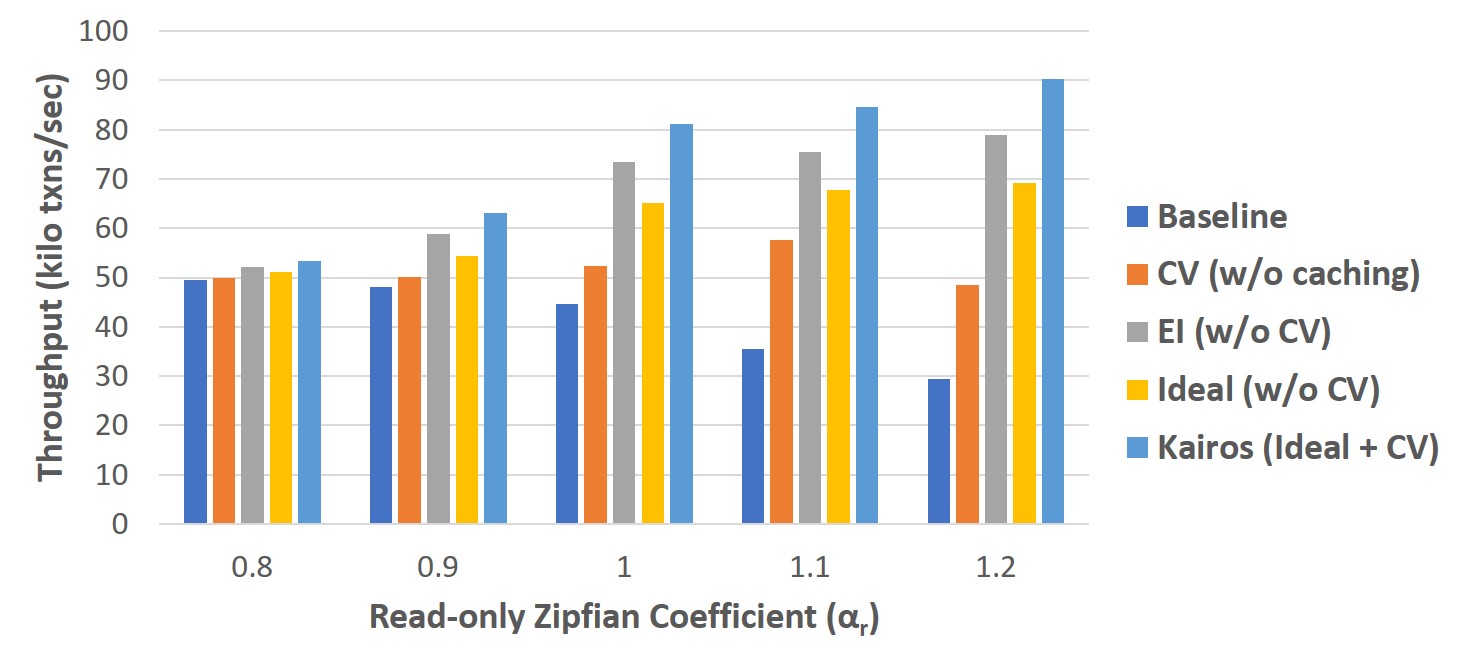} 
\caption{Read-only zipfian coefficient vs throughput}
\label{fig:alpha_r}
\end{subfigure}
\centering
 \begin{subfigure}{0.47\textwidth}
    \centering
 \includegraphics[width=\linewidth,keepaspectratio]{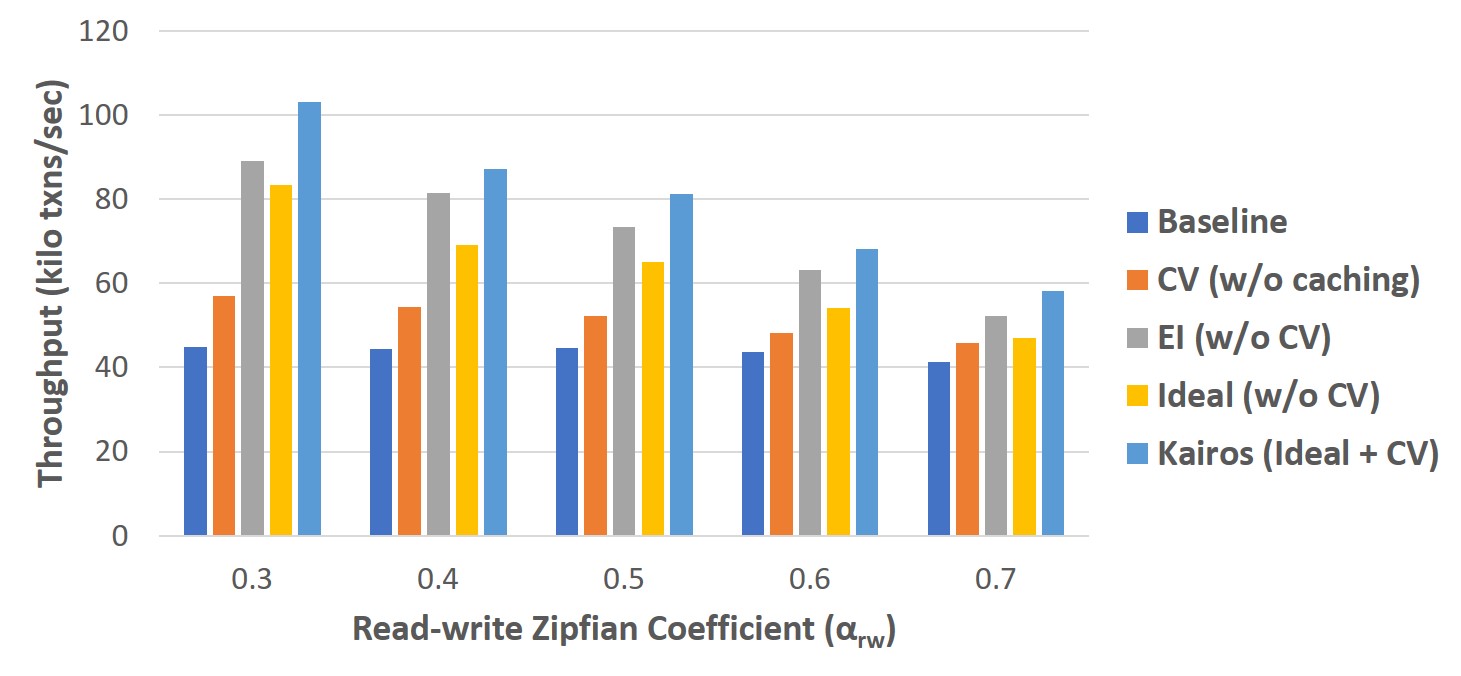}   
\caption{Read-write zipfian coefficient vs throughput}
\label{fig:alpha_rw}
\end{subfigure}
\hfill
\caption{Popularity distribution of keys vs throughput}
\end{figure}

To evaluate the performance impact of our techniques with different popularity distribution of keys, we independently vary the zipfian distribution coefficients that control the popularity of keys accessed in read-only ($\alpha_{r}$) and read-write ($\alpha_{rw}$) transactions, respectively. These coefficients essentially create separate popularity distributions for reading and modifying a key. For example, a key could be read frequently but modified infrequently per the values of the two coefficients. We set the ratio of read-only transactions to 90\% for these experiments.

Figure~\ref{fig:alpha_r} shows the throughput for different values of $\alpha_{r}$; we keep $\alpha_{rw}$ at its default value of 0.5. The skew in the popularity distribution of keys increases with $\alpha_{r}$ and causes a small subset of the keys to be read disproportionately more than others in read-only transactions. In turn, this popularity skew causes the server(s) storing the frequently-read keys to saturate and thereby limit the performance of the entire system. This effect can be seen in the figure. For $\alpha_{r}$ = 0.8, the popularity distribution of keys is less skewed and therefore accesses are more uniformly distributed across the servers. Consequently, techniques like CV and inter-transaction caching do not provide much improvement in performance vs baseline. 

However, as $\alpha_{r}$ increases, the popularity skew increases and therefore the throughput of the baseline suffers because a subset of the servers become a bottleneck and limit the performance of the system. In contrast, techniques like CV and inter-transaction caching provide improvement in performance when the read popularity distribution is skewed because they move the load away from the servers. As seen from the figure, CV alone offers up to 65\% improvement in throughput and inter-transaction caching (EI) alone offers an improvement by up to a factor of 2.7$\times$. Caching provides a higher improvement with increasing $\alpha_r$ since $R_{mean}^{cache}$ of a key keeps decreasing while $W_{mean}^{global}$ remains the same. For example, for a fixed rate of executing transactions, $R_{mean}^{cache}$ of the most frequently read key ranges from 3.2 ms to 160 $\mu$s when $\alpha_{r}$ changes from 0.8 to 1.2, while $W_{mean}^{global}$ remains 160 ms. Consequently, the effectiveness of inter-transaction caching increases since a key can be read more number of times (e.g., from 10 to 44 for the most frequently read key in the workload) from a client's cache after each fetch (miss) from the server. Finally, a combination of the two techniques offers the highest improvement - up to 3.1$\times$ over baseline.

Our next experiment evaluates the impact of varying the write popularity distribution of keys. Figure~\ref{fig:alpha_rw} shows the throughput for different values of $\alpha_{rw}$; we keep $\alpha_{r}$ at its default value of 0.99. As seen from the figure, the performance impact of CV and inter-transaction caching decreases for increasing $\alpha_{rw}$. For a fixed value of $\alpha_{r}$, an increase in $\alpha_{rw}$ causes a frequently-read key to be written more frequently as well. In other words, for a fixed $R_{mean}^{cache}$ for a key, an increase in $\alpha_{rw}$ decreases $W_{mean}^{global}$ for the key and thereby decreases the number of hits from caching the key. For example, for a fixed rate of executing transactions, $W_{mean}^{global}$ of the most frequently read key ranges from 2.8 sec to 10 ms when $\alpha_{rw}$ changes from 0.3 to 0.7, while $R_{mean}^{cache}$ remains 500 $\mu$s causing the number of cache hits after a miss to drop from 100 to 6. Consequently, the effectiveness of client-side caching decreases with an increase in $\alpha_{rw}$. Furthermore, the benefits of CV also decreases with an increase in $\alpha_{rw}$ because the load on servers storing frequently-read keys increases since those keys are written to more frequently.

\subsection{Inter-transaction Caching}
\label{sec:caching_eval}
This section compares the three inter-transaction caching techniques presented in the paper. For all experiments presented in this section, transaction validation occurs on the storage servers. We use up to 30 clients in these experiments, where each client executes 4M transactions with the default workload configuration.

\begin{figure}[t]
\centering
 \includegraphics[width=0.35\textwidth,keepaspectratio]{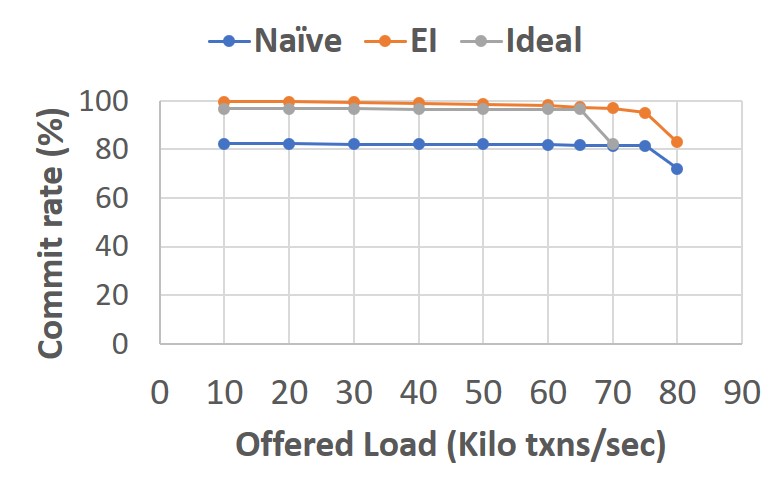}
\caption{Offered load vs transaction commit rate}
\label{fig:load_vs_commit}
\end{figure}

\myparagraph{Impact of offered load on performance} 
This experiment evaluates the impact of increasing offered load on transaction commit rate with the different caching techniques. Figure~\ref{fig:load_vs_commit} shows the results of the experiment. There are a few takeaways from the figure. First, na\"ive caching performs worst as cache management is done only based on transaction decisions with this technique.
Second, EI caching provides up to 3\% higher commit rate than ideal lease-based caching at lower loads but the commit rates wit EI caching drop steadily as load is increased. Increasing load has a greater impact on EI caching because its stale window is impacted by queuing delays due to higher load (see \S\ref{sec:caching_comparison}). Finally, EI caching provides a higher throughput than ideal because ideal has a lower hit rate ($\sim$ 6\%) per client since it relies on probabilistic leases, which causes it to preemptively discard a cached entry to avoid stale hits (vs explicit invalidations). Therefore the system with ideal lease-based caching saturates earlier.

\begin{figure}[tb]
\begin{subfigure}{0.34\textwidth}
    \centering \includegraphics[width=\linewidth,keepaspectratio]{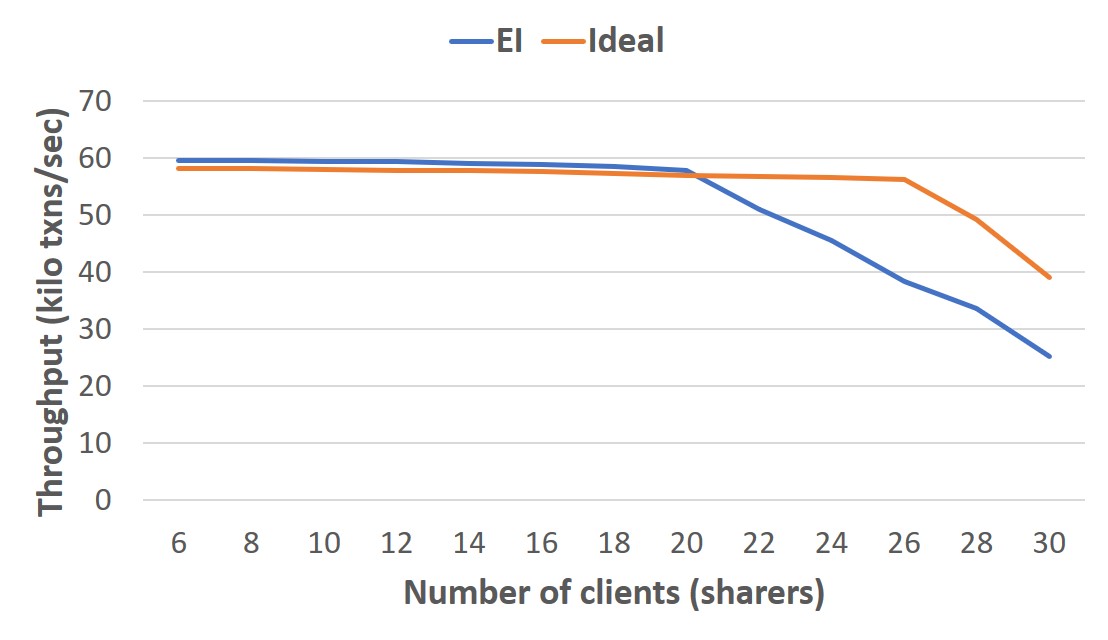} 
\caption{Throughput}
\label{fig:ei_vs_ideal_thr}
\end{subfigure}
\centering
 \begin{subfigure}{0.35\textwidth}
    \centering
 \includegraphics[width=\linewidth,keepaspectratio]{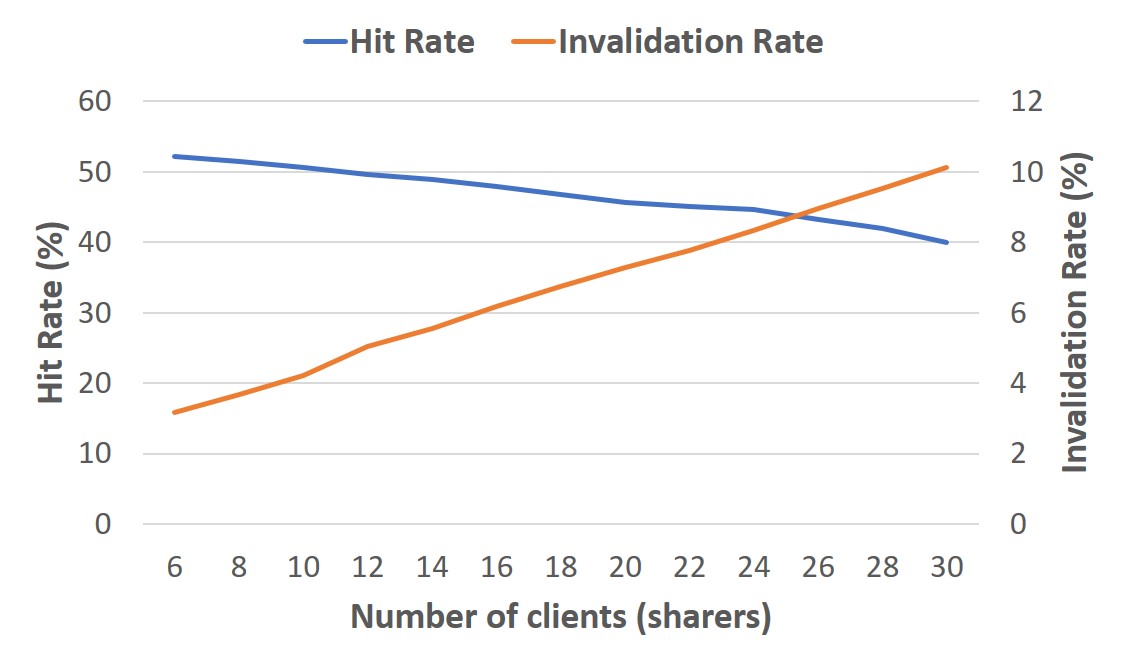}   
\caption{Hit and Invalidation Rate}
\label{fig:ei_vs_ideal_rate}
\end{subfigure}
\hfill
\caption{Scalability with EI and lease-based caching}
\end{figure}

\myparagraph{Impact on scalability}
Here we evaluate the impact of EI and lease-based caching on scalability, while keeping the offered load fixed at 60k transactions/sec (below the saturation point as seen from Figure~\ref{fig:load_vs_commit}).

Figure~\ref{fig:ei_vs_ideal_thr} shows the throughput with each technique for an increasing number of clients. As seen from the figure, the system with EI caching is able to support 20 clients before throughput starts to degrade dramatically. EI caching does not scale well because a server's work for servicing reads and maintaining cache consistency increases linearly with the number of clients and this minimizes the gains from EI caching. Figure~\ref{fig:ei_vs_ideal_rate} shows the hit and invalidation rate per client for an increasing number of clients (note the different y-axis scales for the 2 curves). As seen from the figure, for the same offered load, the hit rate per client decreases (explained later) and the invalidation rate per client increases for increasing number of clients.

In contrast, lease-based caching scales better and is able to support 26 clients (30\% more than EI) before saturation, while offering 46\% higher throughput compared to EI caching. The hit rate per client with lease-based caching also decreases (not shown in figure) with increasing number of clients. This is because, for the same offered load, $R_{mean}^{cache}$ of a key increases while $W_{mean}^{global}$ for the key remains unchanged, thereby reducing the number of hits per cached key. {\em This trend is independent of the caching technique and shows the importance of locality-aware request distribution for effective cache utilization~\cite{pai1998locality,adya2016slicer}}.

Similar to offloading calculation of lease duration to clients, we explored the design space for offloading the work for maintaining client cache consistency with EI from the storage servers to the client (e.g., sharded validator). However, any design almost doubles (worst case) the number of messages in a transaction since all reads (cache misses) during execution would also have to be sent to the client-side cache manager for tracking sharers and writes need to be sent for triggering invalidations.

\subsection{Sensitivity analysis}
Here we evaluate the sensitivity to cache size and lease duration. We use 10 clients and 5 storage shards in all experiments presented in this section. Each client runs the variant YCSB workload with the default configuration (90\% read-only transactions, $\alpha_r$ = 0.99, $\alpha_{rw}$ = 0.5).

\begin{figure}[t]
\centering
 \includegraphics[width=0.35\textwidth,keepaspectratio]{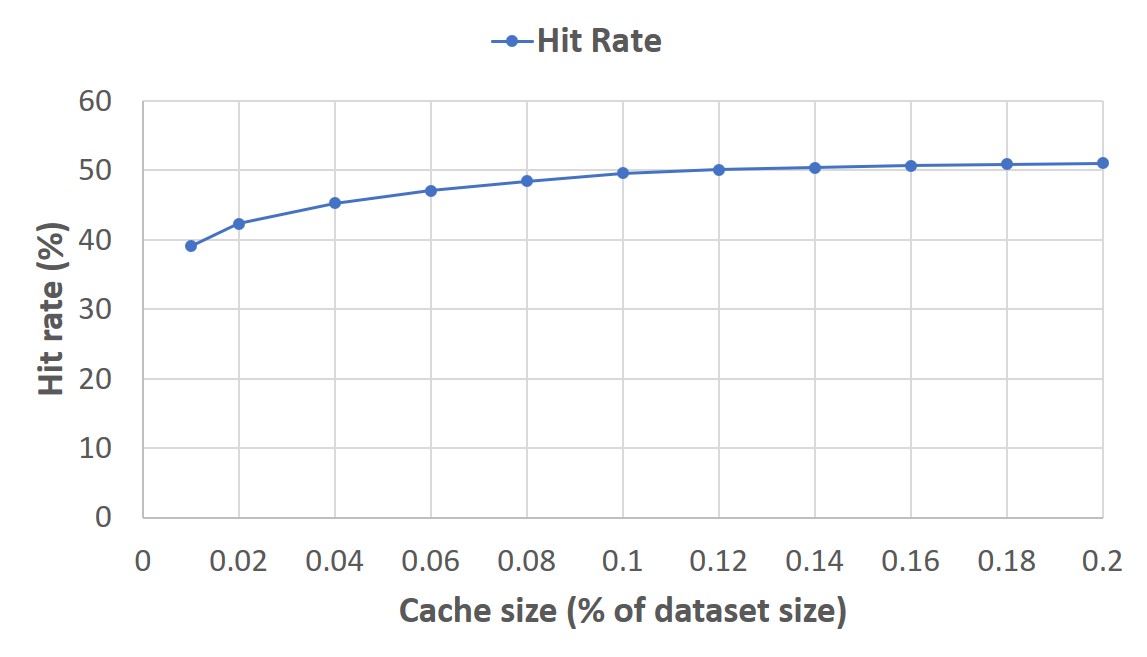}
\caption{Cache size vs hit rate}
\label{fig:size_vs_hit}
\end{figure}

\myparagraph{Impact of cache size} Figure~\ref{fig:size_vs_hit} shows the hit rate for varying cache size (\% of total dataset size). Clients use the na\"ive caching approach; we observe similar trends with other caching techniques. As seen from the figure, initially the hit rate increases rapidly with the cache size, and plateaus after cache size = 0.1\% of the total dataset size. Subsequently, the hit rate only increases by 1.4\% even after doubling the cache size from 0.1\% to 0.2\%. Based on these results, we use a cache size of 0.1\% of the total dataset size in all experiments.

\begin{figure}[t]
\centering
 \includegraphics[width=0.35\textwidth,keepaspectratio]{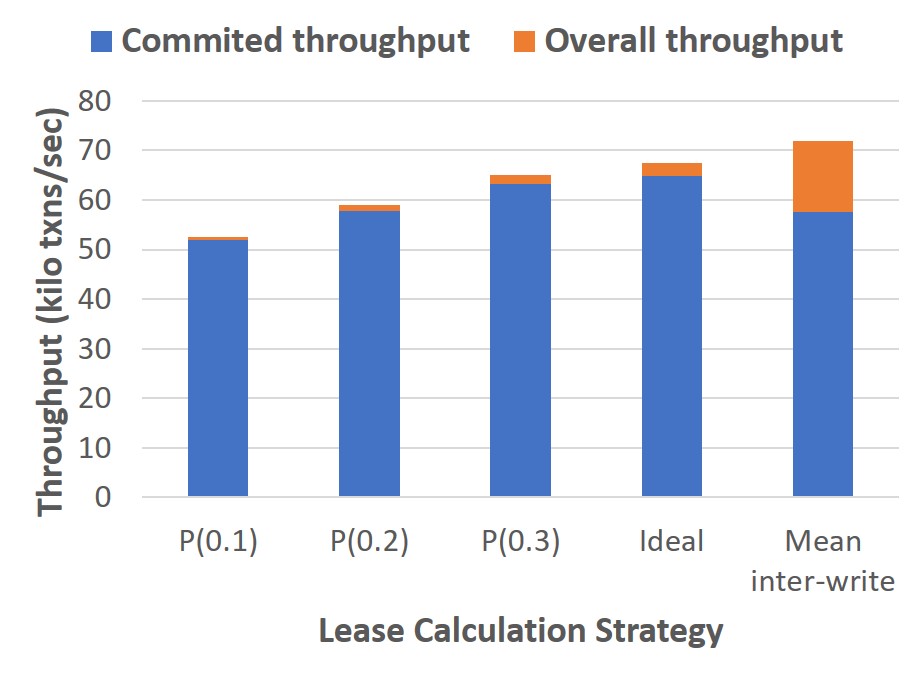}
\caption{Throughput: ideal vs static}
\label{fig:ideal_thr}
\end{figure}

\myparagraph{Impact of using ideal lease duration}
This experiment compares ideal lease calculation with a static approach where leases are based on a fixed probability of an update arriving while a lease is still active. Figure~\ref{fig:ideal_thr} shows the overall and committed throughput with the various lease calculation strategies; the difference between overall and committed indicates the number of aborted transactions per second due to stale cache reads. In the graph, $x$ in $P(x)$ corresponds to the probability of an update arriving within a lease duration $d$, i.e., Pr($W$ $\leq$ $d$) = $x$. The value of $d$ is proportional to $x$, higher values of $x$ lead to higher values of $d$. Note that x in P(x) is not the stale rate. In this experiment, for each key, we find the highest possible value of $d$ that still satisfies the constraint set by the choice of x. The lease duration across all cached keys varies from 4 ms to 5 secs.

As seen from the figure, the committed transaction throughput drops with increasing values of x in P(x). This is because the lease duration increases with the value of x, and the probability of reading a stale value from a client cache increases linearly with the lease duration. Consequently, P(0.4) and mean have higher abort rates. However, simply choosing a lease duration with a lesser probability of reading stale data, e.g., P(0.1), may not be a good strategy either since, even though the commit rate is high, the overall throughput that the system can sustain would be low as the hit rates are lower with shorter leases. Essentially, choosing a lease duration for a key involves a trade-off between hit (impacts overall throughput) and commit rate, and necessitates finding a duration that maximizes hit rate without sacrificing much on commit rate. Our ideal technique is able to achieve this goal through an analytical model (see \S\ref{sec:ideal}). As seen from the figure, it provides the highest committed throughput; the x in P(x) values for ideal lease duration of keys range from 0.04 to 0.5, with more frequently accessed keys having lower values of x.


\section{Related Work}
\label{sec:related}
\paragraph{Caching in distributed storage systems}
Client caching is standard in distributed file systems
~\cite{macklem1994not,pawlowski2000nfs,adya2002farsite} using variants of callback leases~\cite{gray1989leases}.
Prior works have also explored caching to improve performance and/or balance load in key-value stores~\cite{fan2011small,nishtala2013scaling,li2016switchkv,liu2017incbricks,jin2017netcache,vasileios2018cckvs}. However, none of these works support transactions. 


\myparagraph{Client-server transactions with OCC}
Numerous works use optimistic concurrency control (OCC~\cite{kung1981on}) for ACID transactions~\cite{adya1995efficient,dragojevic2014farm,ding2015centiman,zhang2015building,dragojevic2015no,lee2015implementing,chen2016fast,misra2017enabling}.  \sysname follows Thor~\cite{adya1995efficient} in using physical clocks for the OCC version stamps, as do many others (e.g., \cite{zhang2015building,ding2015centiman,misra2017enabling}). Among these, Milana~\cite{misra2017enabling} and Centiman~\cite{ding2015centiman} are most closely related. Milana uses precise clocks for optimizing replication and transaction protocols and also shows that they reduce abort rates in OCC-based systems. However, Milana does not support inter-transaction caching. Comparison with Centiman is presented in \S\ref{sec:comparison_with_centiman}.
Spanner~\cite{corbett2012spanner} also uses physical clocks for transactions, but only for snapshot reads: Spanner does not use OCC.

\myparagraph{Cache consistency} 
Thor~\cite{adya1995efficient} and some of its successors support inter-transaction caching using explicit invalidations (callbacks) to keep caches consistent.  Thor shows that {\it asynchronous} callbacks are sufficient for transaction systems with OCC.  Although asynchrony causes a transaction $T$ to read stale data, consistency is not violated since $T$ fails OCC validation and aborts.  In essence, Thor uses OCC as a fallback for loose cache consistency.  \sysname takes this idea one step further by eliminating the callback entirely (or making it optional).  Thor also shows that OCC with physical clocks leads a rate of spurious aborts that increases with clock skew.  \sysname leverages precise clocks to minimize these aborts (like Milana~\cite{misra2017enabling}) and also to support time-based consistency with a lightweight protocol that directs clients to self-invalidate cached keys at precise times (``soft" leases). 

Sundial~\cite{yu2018sundial} uses leases based on logical time and also integrates inter-transaction caching with OCC for serializable transactions. Sundial is similar to na\"ive caching (see \S\ref{sec:caching_comparison}), but it reorders transactions to avoid some stale hits, and disables caching for a key if its rate of stale hits exceeds an arbitrary configured threshold. 
\sysname uses precise clocks to adjust lease terms dynamically for effective caching on a per-key basis.  Furthermore, \sysname can also support external consistency since transactions commit in physical timestamp order. Finally, \sysname scales validation independent of storage.

\myparagraph{Self invalidation in coherent caches/shared memory systems}
Prior works have used self-invalidation to improve performance of coherence protocols in shared memory multiprocessors~\cite{lebeck1995dynamic,mukherjee1998prediction,lai2000ltp,ross2012complexity}. Mirage~\cite{fleisch1989mirage} uses static leases to reduce coherence overhead in software distributed shared memory systems. 

\section{Conclusion and Future Work}
\label{sec:conclusion}
This paper presents \sysname, a transactional key-value storage system that leverages {\em inter}-transaction caching with self-invalidating leases and sharded validation to alleviate workload-induced hotspots. Our evaluation shows that \sysname offers up to 3.1x the throughput of a baseline system with {\em intra}-transaction caching only. Furthermore, lease-based caching provides better scalability than explicit invalidation-based caching. The limitation of our technique is a small cost of computation and storage to track the inter-access times of keys.

Locality-aware request distribution and precise clocks attract renewed interest in inter-transaction caching, in the context of scalable cloud services. \sysname{} is a first step toward a scalable, lightweight caching infrastructure. Future directions include incorporating the cost of transaction aborts due to stale hits while calculating lease durations and using a bounded soft-state validator like a network switch that is optimized for network I/O.

{\footnotesize \bibliographystyle{acm}
\bibliography{references}}
\balance
\end{document}